# Interaction-Induced Topological Transport of Extended Doublons

*Julius Beck, Matthias Heinrich, Marcus J. Meschede, Helena Drüeke, Francesco S. Piccioli, Sebastian Weidemann, Joshua Feis, Dieter Bauer, and Alexander Szameit**

J. Beck, M. Heinrich, M. J. Meschede, H. Drüeke, F. S. Piccioli, S. Weidemann, J. Feis, D. Bauer, A. Szameit
Institute of Physics, University of Rostock, Albert-Einstein-Str. 23, 18059 Rostock, Germany.
E-mail: alexander.szameit@uni-rostock.de

F. S. Piccioli
INO-CNR BEC Center and Dipartimento di Fisica, Università di Trento, Via Sommarive 14, 38123 Povo, Italy.

**Abstract**

Topological boundary states are a defining feature of topological systems, providing robust transport protected against perturbations. While such states are commonly engineered in non-interacting systems, how to realize interaction-induced topological states composed of only few particles has remained experimentally unexplored thus far. In this work, we realize such states using a driven photonic lattice as a linear analogue platform. We observe the formation of spatially extended bound states of two interacting particles, in which the particles remain correlated while occupying different lattice sites. At discrete values of the local interaction, these bound states support robust transport governed by a topological phase arising from drive-induced correlated hopping. Our results establish an experimentally accessible route to interaction-driven topology beyond single-particle physics and provide a versatile framework for exploring correlated few-particle phenomena across different physical platforms.

## 1. Introduction

Shortly after their discovery [1,2] and subsequent first experimental demonstration [3], topological insulators began to attract substantial interest by virtue of their peculiar properties, in particular their signature feature of scattering-immune protected edge currents [4]. Today, topological insulators, and topological transport in general, are no longer exclusive to condensed-matter systems [5], but have become a staple in numerous other wave-physical systems, such as cold atoms [6,7], electric circuits [8,9], mechanical waves [10], acoustics [11], and photonics [12,13]. This wide variety of available platforms provides exceptional conditions for different methods of establishing non-trivial topology to be explored and honed, such as anomalous driving [14–16], non-Hermitian topology [17,18], quantum topology [19,20], adiabatic Thouless pumping [21–23], and nonlinear topological systems [9,24–26].

Nonlinear systems play a central role in the understanding of many-body topological phenomena as interaction forces between many particles can be captured by nonlinear terms using mean-field approaches. The prototypical example is the Gross-Pitaevskii equation [27], where a nonlinear Kerr term is added to a single-particle Schrödinger equation in order to describe the average dynamics resulting from collisions [9,25,28]. At the same time, inter-acting few-particle systems offer a complementary route that bypasses mean-field nonlinearities and instead probes genuine two-particle correlations inherent to the linear quantum many-body problem. These correlated dynamics can be faithfully reproduced in linear analogue platforms using engineered couplings or dimensional mappings [29,30].

On this few-particle level, particular attention has been given to bipartite quasiparticles known as Doublons [29]. These metastable states consist of two particles bound by either repulsive [29] or attractive interactions [31,32] and exhibit remarkably long lifetimes even when experiencing intra-state collisions [29]. The existence of Doublons also facilitates many-body localization in disordered [31] and generalized [33] Hubbard models. Topologically protected Doublon boundary states may exist at parameter regimes where single particle systems are topologically trivial [34,35], and sufficiently strong interactions may even break the conventional bulk-boundary correspondence [34,35], requiring its extension to interacting two-particle systems [36]. Moreover, topological Doublons can be interpreted as generalized W states of qutrit systems, offering a platform to study many body quantum correlations [37].

Previous experimental realizations, ranging from cold-atom to photonic systems [29,31,32,38–42], have primarily focused on localized Doublons – states in which both particles occupy the same lattice site. Yet, bound states of interacting particles do not necessarily require double occupancy. In particular, they may also form spatially separated correlated pairs, as discussed in a few theoretical studies on adjacent-site and extended Hubbard configurations [43,44]. We refer to such states as extended Doublons – bound two-particle states in which the particles occupy different lattice sites while moving coherently as a correlated unit.

In this work, we experimentally explore the physics of extended Doublon quasi-particles in a two-dimensional driven lattice. The formation of the Doublons themselves arises from the local on-site Hubbard interactions, binding the two particles into a metastable state. Their non-trivial topology, however, arises from drive-induced correlated hopping processes enabled by these local interactions. Because our system operates in a fully linear photonic regime, the dimensional mapping between the driven two-dimensional lattice and the two-particle Hubbard model provides a direct correspondence to the correlated dynamics of two interacting particles. In contrast to previous realizations where topological behavior was engineered through single-particle band structures, here the topological phase originates from the interplay of local interactions and dynamically generated correlated hopping.

By employing a dimensional mapping, we directly observe the doublon dynamics and their topologically protected transport. Using a photonic artificial solid as a model platform, we demonstrate that these Doublons form at discrete resonances of the on-site interaction strength and propagate as interaction-induced topological edge states through the bulk of the system. This establishes a new class of interaction-driven, yet linearly realizable, topological transport bridging non-interacting topological systems and correlated quasiparticle dynamics.

## 2. Results

### 2.1. Concept and mapping

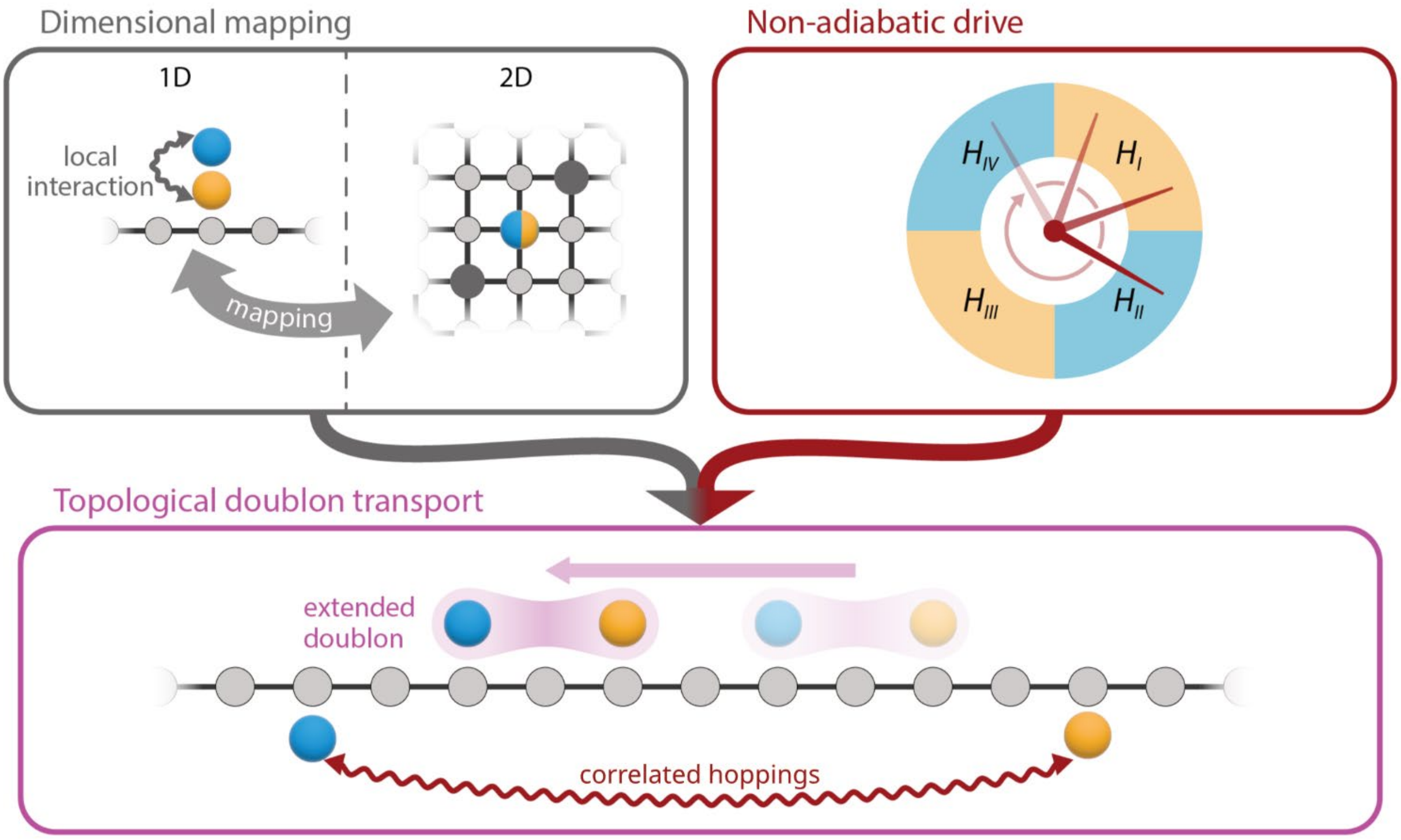

**Figure 1. Topological bulk transport of extended Doublons.** Dimensional mapping enables the observation of two-particle dynamics with local interactions in single-particle systems. The local interactions appear as onsite potentials along the diagonal in the 2D lattice. When the couplings of the underlying 1D chain are activated sequentially by a non-adiabatic periodic drive (see Supplementary Information), correlated hopping processes between the particles become dynamically active, giving rise to a non-trivial topological phase. The combination of dimensional mapping and non-adiabatic driving enables the experimental observation of the topological bulk transport of an extended two-particle bound state (extended Doublon).

Our approach is summarized in **Figure 1**. We start with a Su-Schrieffer-Heeger chain[45,46] characterized by alternating weak and strong bonds between the individual lattice sites, and place two distinguishable particles that exhibit a local (i.e., hard-shell) interaction. This system is governed by the Hamiltonian

$$\hat{H} = \hat{H}_0 + \hat{H}_{int} , \quad (1)$$

where $\hat{H}_0$ describes the individual dynamics of each particle under nearest-neighbor hopping, and $\hat{H}_{int}$ accounts for their local interaction. As explained in detail in the Supplementary Information, $\hat{H}_0$ can be expanded into four virtual Hamiltonians $\hat{H}_I, \dots, \hat{H}_{IV}$,

$$\hat{H}_0 = \hat{H}_I + \hat{H}_{II} + \hat{H}_{III} + \hat{H}_{IV}, \quad (2)$$

that, individually, consist exclusively of correlated hopping terms. In the static system, these correlated hopping contributions cancel exactly in the full Hamiltonian $\hat{H}_0$. Crucially, this mutual cancellation in the summation only occurs if the four virtual Hamiltonians act simultaneously. The situation changes dramatically if the system is subjected to a time-dependent driving protocol that sequentially activates the constituent Hamiltonians. By preventing their mutual cancellation, the non-adiabatic drive renders the correlated hopping process dynamically active and directly observable in the particles' motion. To this end, our non-adiabatic drive establishes different delays on the individual virtual Hamiltonians. As we will show in the following, the drive-induced correlated hopping enables the formation of extended doublon states, whose collective dynamics exhibits a topological character. As was shown before[30,38], the motion of two interacting particles on a one-dimensional (1D) lattice can be mapped onto the evolution of a single particle in a two-dimensional (2D) square lattice. In this mapping, the position coordinates of the two particles in the 1D lattice correspond to the two Cartesian coordinates of the 2D lattice. Accordingly, the term "particle" refers here to the synthetic coordinates of the mapped two-particle wavefunction rather than to physical particles. The local Hubbard interaction strength in the 1D system is represented by an on-site detuning of the diagonal sites of the 2D system (**Figure 2a**).

### 2.2. Driving protocol and the emergence of correlated dynamics

The prerequisite non-adiabatic drive is imposed by means of the protocol depicted in Figure 2b. While the on-site potential remains constant over the entire Floquet period $T$, coupling to each of a lattice site's four neighbors is selectively enabled in one of the four discrete steps for $T/4$ in clockwise order. As detailed in the Supplementary Information, this drive preserves the validity of the dimensional mapping and, moreover, each step of the protocol directly instantiates one of the virtual Hamiltonians $\hat{H}_I, \dots, \hat{H}_{IV}$. Whereas in the static system, the correlated hopping processes is absent due to their cancellation, the driven sequence reveals their physical impact on the two-particle dynamics. In each step, the position of one particle determines which hopping processes are available to the other, giving rise to a correlated motion that is absent in the undriven lattice.

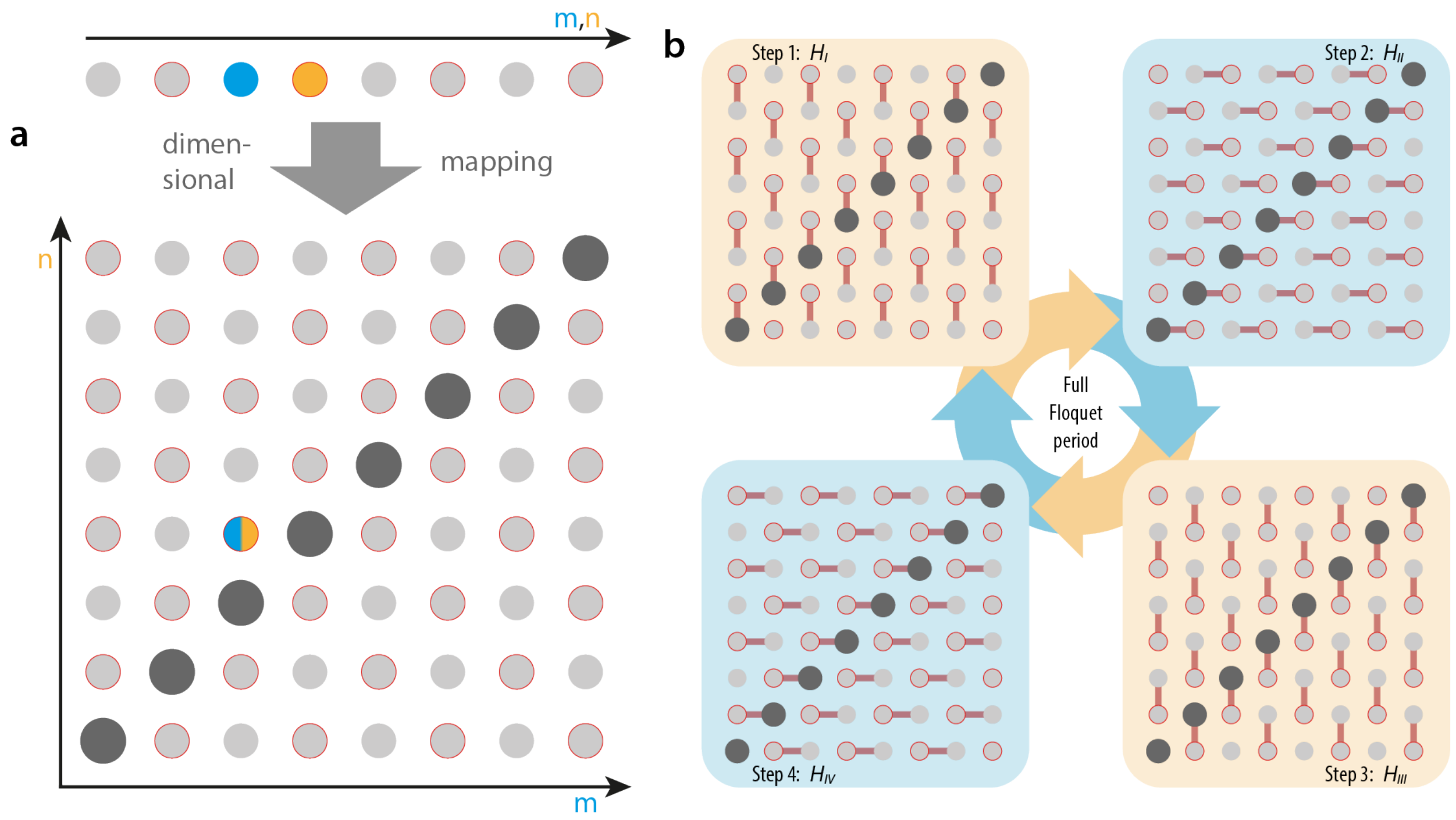

**Figure 2. Dimensional mapping. (a)** The dynamics of two interacting particle (orange and blue) with positions $m$ and $n$ on a 1D lattice can be mapped to a single particle on a corresponding square lattice. There the local on-site interaction is governed by a detuning of the diagonal lattice sites (highlighted in dark grey), corresponding to a Hubbard interaction. **(b)** The system is externally driven by applying a time periodic modulation to the couplings between both sublattices (indicated by red and grey outline) of the 2D lattice, according to the anomalous Floquet protocol. The protocol consists of four discrete couplings steps lasting for $T/4$ of full period $T$, while the onsite potential of the sites stays constant over a full cycle. Each step realizes one of the virtual Hamiltonians $\hat{H}_I, \ldots, \hat{H}_{IV}$, and their sequential activation prevents mutual cancellation of correlated hopping contributions.

## 2.3. Topological structure and resonant decoupling

A perhaps more widely known consequence of the helical drive is its analogue to time-reversal symmetry-breaking that renders the lattice topologically non-trivial as indicated by non-vanishing winding numbers[47]. Along these lines, recent implementations of anomalous topological insulators[14–16], $\mathbb{Z}_2$ topological insulators[48], and even PT-symmetric topological insulators[17] have been reported.

In the context of our mapping procedure, we note that anomalous topological edge states persist in the presence of local perturbations such as detuned diagonal sites of the square lattice. Moreover, since the topological characteristics of a lattice are unaffected by its shape, a triangular lattice domain subjected to the same drive equally supports topological edge states along its perimeter. This situation corresponds to the limit of an infinitely detuned diagonal, which effectively bisects the square lattice into two uncoupled triangular domains by fully suppressing hopping across the diagonal.

In this limit, topological edge states propagate along channels directly adjacent to the detuned diagonal. Viewed through the lens of the mapping procedure, these states correspond to a bound two-particle configurations in which both particle coordinates increase or decrease together, forming an extended doublon that traverses the bulk of the effective 1D system. Notably, the sign of the interaction does not play a role: both positive and negative detunings (corresponding to repulsive and attractive forces) give rise to such states.

## 2.4. Finite interaction strengths and resonances

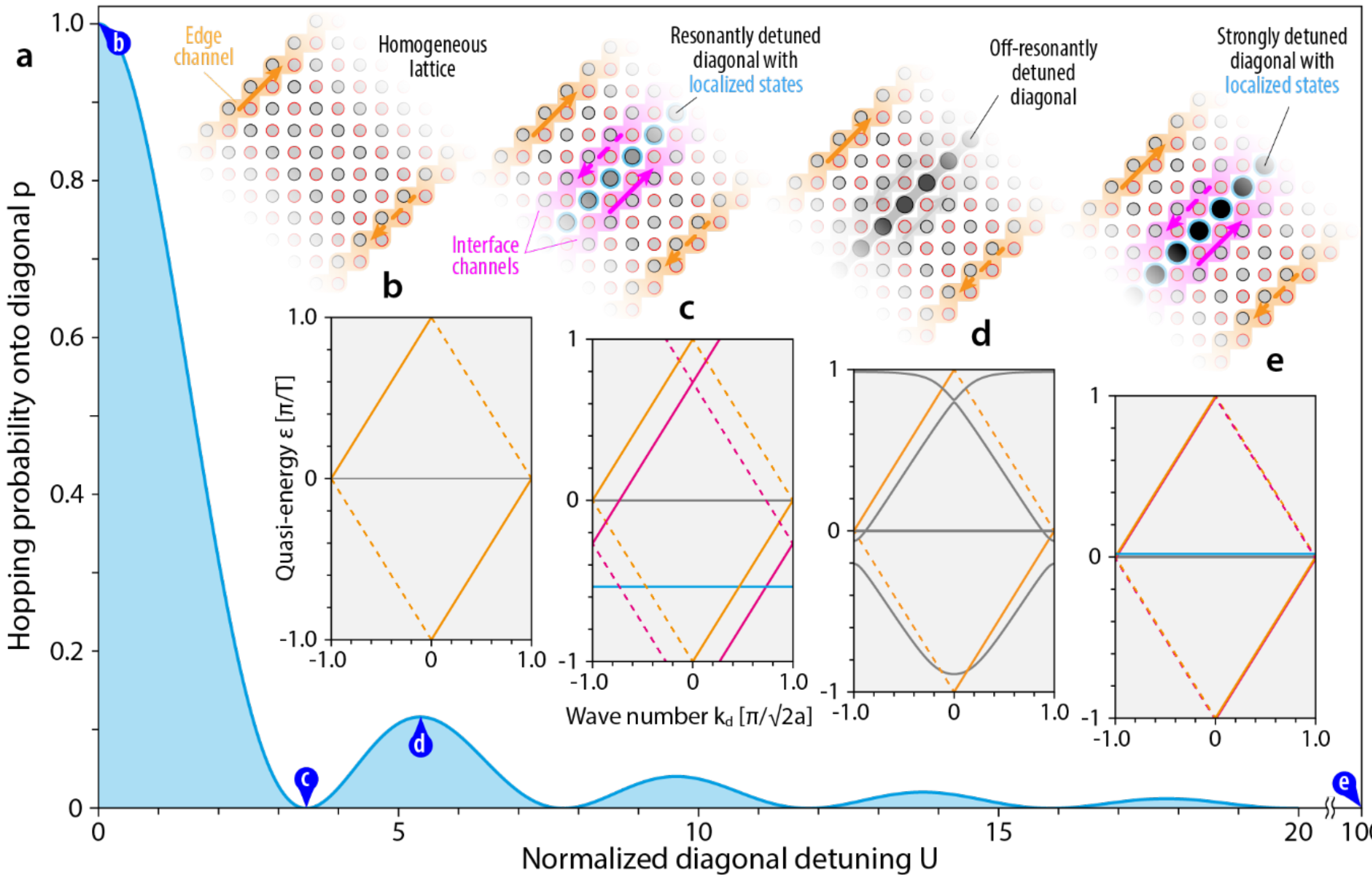

**Figure 3. Transfer probability and band structures. (a)** The hopping probability $p(U)$ onto the lattice's diagonal as a function of the interaction strength $U$, showing an oscillatory behavior with vanishing values at specific resonaces. **(b)** For zero diagonal detuning ($U = 0$), corresponding to vanishing interaction strength, an anomalous edge state (orange) bifurcates from the degenerate flat-bands at zero energy. **(c)** At the first zero of $p(U)$ the lattice domains above and below the diagonal decouple, giving rise to additional interface channels with corresponding edge bands (pink). The states on the diagonal sites remain localized in the detuned flat-band (blue). **(d)** For non-zero detuning off the resonances, a superposition of three interface bands with avoided crossings forms. **(e)** When increasing the detuning towards infinity the (pink) interface bands and the conventional edge bands (orange) overlap in energy.

Beyond this transparent limiting case, the infinite-interaction regime represents only a special example of a more general phenomenon. Topological bulk states persist for finite interaction strengths[37,49]. **Figure 3a** shows the hopping probability $p(U)$ onto the lattice diagonal as a function of the detuning $U$ of the diagonal sites. While its envelope decays toward zero for infinitely strong detunings, $p(U)$ exhibits an oscillatory behavior and vanishes for specific values of $U$.

At these resonances, the two triangular lattices are again effectively decoupled, supporting interface channels with topological character. This interpretation is supported by the band structure of our system depicted in Figure 3b-d. In the trivial case without detuning on the diagonal ($p(0) = 1$) an anomalous edge state bifurcates from the degenerate flat bands at zero energy. At the first zero of the hopping probability ($p(2\sqrt{3}) = 0$) the lattices above and below the detuned diagonal are decoupled, resulting in additional interface channels with corresponding edge bands. In turn, the states on the diagonal sites remain localized in the detuned flat band. For off-resonant detunings ($p(U) > 0$), a superposition of three interface bands displaying avoided crossings are present until the interface bands and the conventional edge bands eventually overlap in energy at $U \to \infty$.

## 2.5. Experimental observation of extended doublon dynamics

For our experiments, we implemented our driven lattices on a photonic platform: Evanescently coupled integrated photonic waveguides fabricated via the femtosecond laser direct-write technique[50] (see Methods). The evolution of light in waveguide lattices is governed by a set of tight-binding equations that are analogous to the discrete Schrödinger equations known from condensed matter physics. The main distinction, however, is that the time coordinate $t$ is replaced by the spatial propagation coordinate $z$ along the waveguides in the photonic system. The non-adiabatic drive, hence, can be established through a modulation of the inter-site hopping along the waveguide structure in a chiral fashion that breaks the $z$-reversal symmetry of the system.

In **Figure 4**, we present a summary of our measurements of the Doublon dynamics with finite interaction strength at the first resonance. The system was probed incrementally by single-site excitations along the off-diagonal of the square lattice of waveguides at different positions in each panel. After two full Floquet periods, the wave packet has propagated over four sites, that is, two unit cells, in clockwise direction. In each step of the Floquet protocol, one particle remains effectively stationary and thereby governs the motion of the other, realizing correlated hopping processes that are independent of the particles' separation.

This dynamics constitutes the hallmark of an extended doublon: one particle moves only when the other occupies a specific neighboring configuration, and vice versa. The resulting motion is of topological origin, as the excitation populates an anomalous edge mode at the detuned diagonal of the mapped 2D lattice.

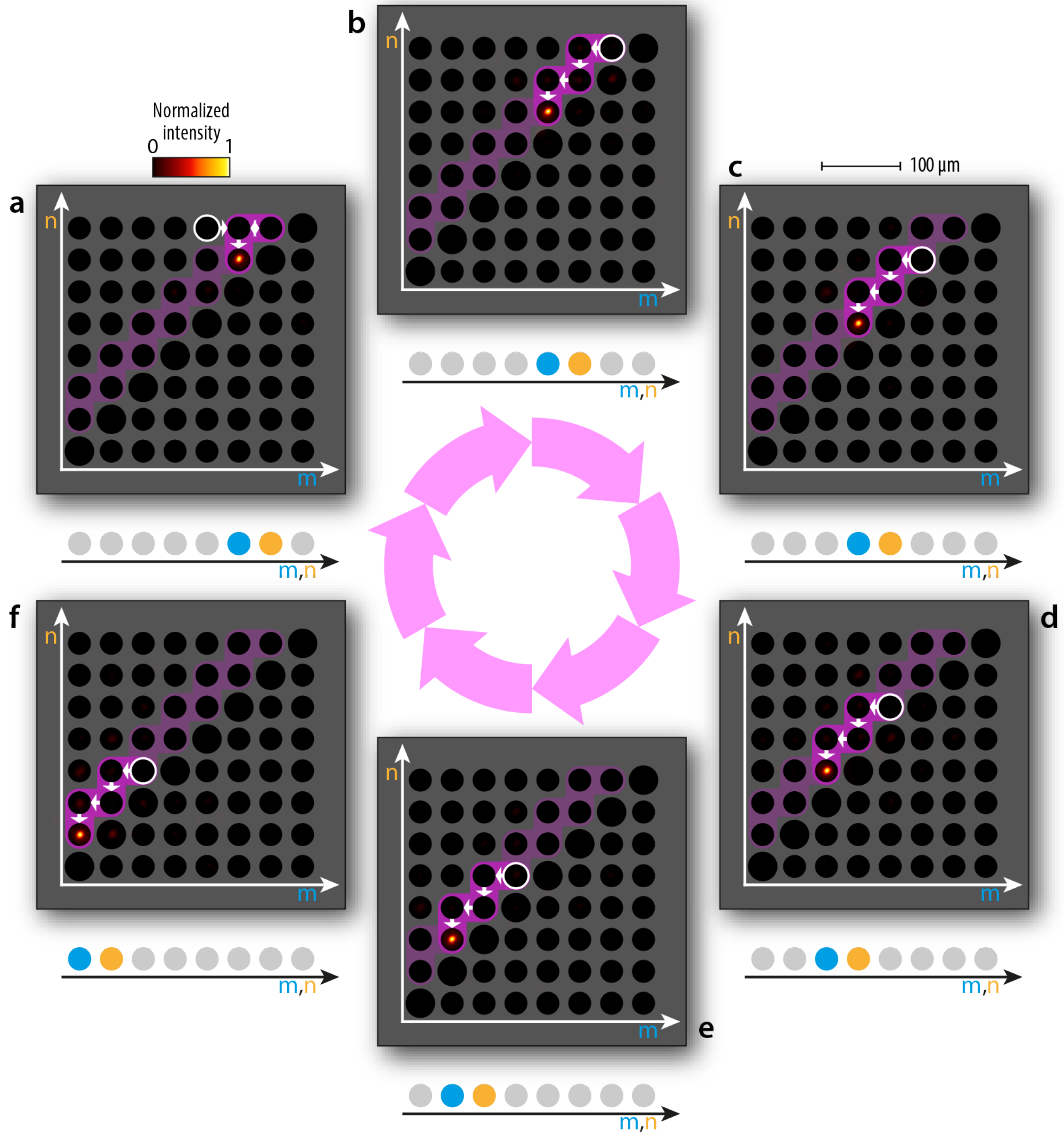

**Figure 4. Tracing the extended-Doublon state at the first resonance**. The resonant system was probed experimentally by successive single-site excitations (white circles) along the off-diagonal of the square lattice. By setting the excitation to the output of the previous measurement panel, the movement of the edge state can be traced. (**a-f**) After two full Floquet periods the injected wave packet had propagated over four sites (i.e. two unit cells) along channels adjacent to the main diagonal. When mapped back onto the one-dimensional lattice, this motion corresponds to the propagation of an extended doublon trough the bulk. The bound state remains spatially extended over two sites throughout the propagation. Assuming periodic boundary conditions in the one-dimensional lattice, panel (**f**) would follow another instance of panel (**a**).

A comparison of the Doublon dynamics at different interaction strengths, that is, different detunings of the diagonal, is provided in **Figure 5**. The panels on the left show the dynamics for zero interaction. In this case, Doublons cannot form and no joint dynamics through the bulk are observed. Increasing the detuning beyond the first resonance leads to unstable extended Doublon states, in which the excitation remains confined to the diagonal and adjacent doublon channels but gradually redistributes between them, resulting in a loss of coherent propagation. In the limit of very strong interactions, the dynamics converge to the resonant behavior.

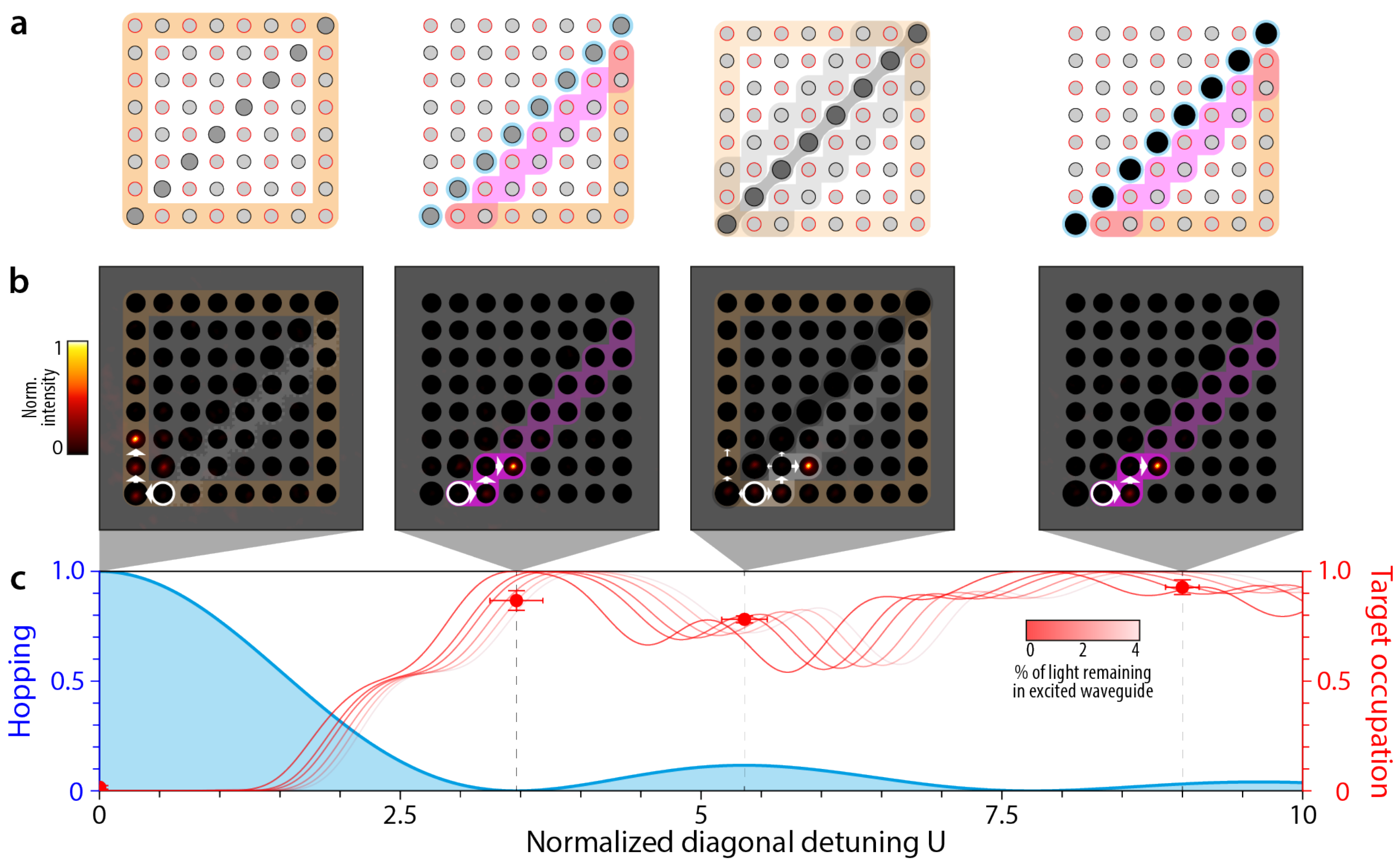

**Figure 5. Comparison of the dynamics for different interaction strengths**. **(a)**: Schematic illustration of the available propagation channels from the excitation site. The compact localized states are indicated in blue, conventional topologically protected states in orange, and the extended-Doublon edge channel in pink. Meanwhile, non-topological channels are distinguished in grey. **(b)** Snapshot of the experimentally measured light propagation after two Floquet periods for various particle interaction potentials, when the white circled waveguide was populated. The light path is emphasized by arrows, where the line thickness is an indication of the coupling probability. **(c)** Illustration of the hopping probability onto the main diagonal (blue) and target occupation of the extended-Doublon channel as a function of interaction strength. Solid curves correspond to numerical simulations, while discrete data points with error bars represent experimental measurements at the four interaction strengths investigated. All values were evaluated using the same target-occupation definition (see Supplementary Information). The solid theoretical curves of the target ratio emphasize the fluctuation range of the manufacturing accuracy of the targeted 100:0% directional couplers [14].

Finaly, experimental data probing the characteristics bands at the first resonance (see Figure 3c) are shown in Figure S1, including the dispersionless bulk flat band, the diagonal doublon flat band, and the interface edge modes.

## 3. Conclusion

Our results demonstrate a distinct route toward realizing topological transport in interacting few-particle systems. In contrast to conventional approaches, where topology is inherited from single-particle band structures or described within mean-field nonlinear frameworks, the topological phase observed here emerges in the two-particle configuration space and is encoded in correlated kinetic processes. Local on-site interactions bind the particles into composite quasiparticles, while a non-adiabatic drive generates correlated hopping processes that give rise to topological transport. This clear separation of roles allows topology to arise from particle correlations rather than from static band geometry.

Within this framework, we experimentally observe the formation of extended doublons - bound two-particle states in which the particles remain spatially separated while moving coherently. At discrete values of the local interaction strength, these states propagate robustly through the bulk of the system as topological edge states. Using a photonic artificial solid in combination with dimensional mapping to a two-particle Hubbard model, we directly access and visualize this interaction-enabled transport in a fully linear platform.

Beyond the specific photonic implementation, the mechanism demonstrated here is broadly applicable to a wide class of synthetic and quantum systems in which time-dependent control and interactions can be combined. Our results open new experimental pathways for studying interaction-driven topology in few-particle settings, bridging the gap between non-interacting topological systems and many-body correlated phases.

## 4. Materials and Methods

### 4.1. Experimental configuration

The photonic structures were fabricated using the femtosecond laser direct-writing method[50]. Ultrashort laser pulses emitted from a frequency-doubled fiber amplifier system (Coherent Monaco, wavelength $517$ nm, repetition rate $333$ kHz, pulse duration $270$ fs) were focused into the volume of a fused silica sample (Corning 7980, dimensions $1\ \text{mm} \times 20\ \text{mm} \times 150\ \text{mm}$, bulk refractive index $n_0 = 1.457$ at $633$ nm), utilizing a 50x microscope (0.60 NA) objective. In the focal region, a permanent refractive index change is induced and, by moving the sample along arbitrary three-dimensional trajectories under the beam via a precision translation system (Aerotech ALS130), waveguides are formed. The resulting single-mode waveguides have a typical refractive index contrast of up to $\Delta n_0 = 2 \cdot 10^{-3}$. The local on-site interactions, represented by detunded waveguides along the diagonal of the 2D square lattice, were realized by inscribing these waveguides at different speeds between 35mm/min and 120mm/min. Lower writing speeds correspond to higher refractive index contrast and thus to larger effective on-site interaction potentials.

The drive-induced correlated hopping processes were implemented via a non-adiabatic anomalous driving protocol[14,47] over two full Floquet periods $T$, by selectivity activating nearest-neighbor couplings using directional couplers. To achieve a non-trivial winding of $W = 1$, a coupling strength of $J = 2\pi/T$ is chosen, corresponding to a complete light transfer in each coupler. In the experimental realization, this results in a coupling strength of $J = 0.1189\ \text{mm}^{-1}$ for couplers of the length $L = 13.211$ mm.

In practice, the realized coupling ratios deviate slightly from the ideal 100:0 transfer, with up to 4% of optical power remaining in the initially excited waveguide. These deviations were explicitly included in the numerical simulations used for comparison with the experimental data, as shown in Figure 5c. The normalized interactions strengths $U$ were defined as the ratio of the absolute detuning $\delta$ (extracted from an independent detuning calibration scan) to the coupling strength $J$.

The propagation dynamics of the Doublon were probed with coherent light from a Helium-Neon laser (Melles Griot - 25mW), and measured after two Floquet periods using a CCD camera (Basler A120f).

### 4.2. Numerical simulations

The band structures (Figure 2) are calculated numerically by solving the discrete Schrödinger equation as an eigenvalue problem. A finite grid strip (as shown in Figure 2) with periodic boundary conditions in the direction of the wave vector was used for this purpose.

The target occupation shown in Figure 5c was obtained from numerical simulations of the discrete Schrödinger dynamics and is defined as the fraction of the total optical intensity occupying the targeted extended-doublon channel relative to the total intensity in the system. The continuous curves correspond to numerically simulated target occupations as a function of the normalized interaction strength, while the discrete data points represent experimentally measured values at the four interaction strengths investigated. In both cases, the target occupation was evaluated using the same definition and analysis procedure, with experimental error bars reflecting statistical uncertainties of the measured intensity distributions. Further details on its computation and interpretation are provided in the Supplementary Information.

### Acknowledgements

We thank C. Otto for preparing the high-quality fused silica samples used for the inscription of all photonic structures employed in this work. A.S. acknowledges funding from the Deutsche Forschungsgemeinschaft (grants SZ 276/9-2, SZ 276/19-1, SZ 276/20-1, SZ 276/21-1, SZ 276/27-1, and GRK 2676/1-2023 'Imaging of Quantum Systems', project no. 437567992). A.S. also acknowledges funding from the Krupp von Bohlen and Halbach Foundation as well as from the FET Open Grant EPIQUS (grant no. 899368) within the framework of the European H2020 pro-gramme for Excellent Science. A.S., M.H. and D.B. acknowledge funding from the Deutsche Forschungsgemeinschaft via SFB 1477 'Light–Matter Interactions at Interfaces' (project no. 441234705).

### Author contributions

J.B. designed and fabricated the samples and performed the experiments. M.M., H.D., D.B. provided the theoretical foundations. J.B., F.P. performed the numerical simulations. S.W., J.F., M.H. provided analysis tools. A.S. supervised the project. All authors discussed the results and co-wrote the manuscript.

### Competing interests

The authors declare that they have no competing interests.

**Data Availability Statement**

The experimental and numerical data that support the findings of this study are available in an open-access data repository [reference follows].

## References

1. Kane, C. L. & Mele, E. J. Quantum Spin Hall Effect in Graphene. Phys. Rev. Lett. 95, 226801 (2005).

2. Bernevig, B. A. & Zhang, S.-C. Quantum Spin Hall Effect. Phys. Rev. Lett. 96, 106802 (2006).

3. König, M. et al. Quantum Spin Hall Insulator State in HgTe Quantum Wells. Science 318, 766–770 (2007).

4. Hasan, M. Z. & Kane, C. L. Colloquium : Topological insulators. Rev. Mod. Phys. 82, 3045–3067 (2010).

5. Ozawa, T. et al. Topological photonics. Rev. Mod. Phys. 91, 015006 (2019).

6. Haldane, F. D. M. Model for a Quantum Hall Effect without Landau Levels: Condensed-Matter Realization of the 'Parity Anomaly'. Phys. Rev. Lett. 61, 2015–2018 (1988).

7. Jotzu, G. et al. Experimental realization of the topological Haldane model with ultracold fermions. Nature 515, 237–240 (2014).

8. Imhof, S. et al. Topolectrical-circuit realization of topological corner modes. Nature Phys 14, 925–929 (2018).

9. Hadad, Y., Soric, J. C., Khanikaev, A. B. & Alù, A. Self-induced topological protection in nonlinear circuit arrays. Nat Electron 1, 178–182 (2018).

10. Süsstrunk, R. & Huber, S. D. Observation of phononic helical edge states in a mechanical topological insulator. Science 349, 47–50 (2015).

11. He, C. et al. Acoustic topological insulator and robust one-way sound transport. Nature Phys 12, 1124–1129 (2016).

12. Rechtsman, M. C. et al. Photonic Floquet topological insulators. Nature 496, 196–200 (2013).

13. Hafezi, M., Mittal, S., Fan, J., Migdall, A. & Taylor, J. M. Imaging topological edge states in silicon photonics. Nature Photon 7, 1001–1005 (2013).

14. Maczewsky, L. J., Zeuner, J. M., Nolte, S. & Szameit, A. Observation of photonic anomalous Floquet topological insulators. Nat Commun 8, 13756 (2017).

15. Mukherjee, S. et al. Experimental observation of anomalous topological edge modes in a slowly driven photonic lattice. Nat Commun 8, 13918 (2017).

16. Wintersperger, K. et al. Realization of an anomalous Floquet topological system with ultracold atoms. Nat. Phys. 16, 1058–1063 (2020).

17. Fritzsche, A. et al. Parity–time-symmetric photonic topological insulator. Nat. Mater. 1–6 (2024) doi:10.1038/s41563-023-01773-0.

18. Poli, C., Bellec, M., Kuhl, U., Mortessagne, F. & Schomerus, H. Selective enhancement of topologically induced interface states in a dielectric resonator chain. Nat Commun 6, 6710 (2015).

19. Blanco-Redondo, A., Bell, B., Oren, D., Eggleton, B. J. & Segev, M. Topological protection of biphoton states. Science 362, 568–571 (2018).

20. Mittal, S., Goldschmidt, E. A. & Hafezi, M. A topological source of quantum light. Nature 561, 502–506 (2018).

21. Kraus, Y. E., Lahini, Y., Ringel, Z., Verbin, M. & Zilberberg, O. Topological States and Adiabatic Pumping in Quasicrystals. Phys. Rev. Lett. 109, 106402 (2012).

22. Zilberberg, O. et al. Photonic topological boundary pumping as a probe of 4D quantum Hall physics. Nature 553, 59–62 (2018).

23. Lohse, M., Schweizer, C., Price, H. M., Zilberberg, O. & Bloch, I. Exploring 4D quantum Hall physics with a 2D topological charge pump. Nature 553, 55–58 (2018).

24. Mukherjee, S. & Rechtsman, M. C. Observation of Floquet solitons in a topological bandgap. Science 368, 856–859 (2020).

25. Maczewsky, L. J. et al. Nonlinearity-induced photonic topological insulator. Science 370, 701–704 (2020).

26. Jürgensen, M., Mukherjee, S., Jörg, C. & Rechtsman, M. C. Quantized fractional Thouless pumping of solitons. Nat. Phys. 19, 420–426 (2023).

27. Pitaevskii, Lev P. Pitaevskii, Lev P. 'Vortex lines in an imperfect Bose gas.' Sov. Phys. JETP 13.2 (1961): 451-454.

28. Bloch, I. Ultracold quantum gases in optical lattices. Nature Phys 1, 23–30 (2005).

29. Winkler, K. et al. Repulsively bound atom pairs in an optical lattice. Nature 441, 853–856 (2006).

30. Longhi, S. Photonic Bloch oscillations of correlated particles. Opt. Lett., OL 36, 3248–3250 (2011).

31. Schreiber, M. et al. Observation of many-body localization of interacting fermions in a quasirandom optical lattice. Science 349, 842–845 (2015).

32. Walter, A.-S. et al. Quantization and its breakdown in a Hubbard–Thouless pump. Nat. Phys. 19, 1471–1475 (2023).

33. Barbiero, L., Menotti, C., Recati, A. & Santos, L. Out-of-equilibrium states and quasi-many-body localization in polar lattice gases. Phys. Rev. B 92, 180406 (2015).

34. Di Liberto, M., Recati, A., Carusotto, I. & Menotti, C. Two-body physics in the Su-Schrieffer-Heeger model. Phys. Rev. A 94, 062704 (2016).

35. Gorlach, M. A. & Poddubny, A. N. Topological edge states of bound photon pairs. Phys. Rev. A 95, 053866 (2017).

36. Stepanenko, A. A. & Gorlach, M. A. Interaction-induced topological states of photon pairs. Phys. Rev. A 102, 013510 (2020).

37. Berti, A. & Carusotto, I. Topological two-particle dynamics in a periodically driven lattice model with on-site interactions. Phys. Rev. A 105, 023329 (2022).

38. Corrielli, G., Crespi, A., Della Valle, G., Longhi, S. & Osellame, R. Fractional Bloch oscillations in photonic lattices. Nat Commun 4, 1555 (2013).

39. Mukherjee, S. et al. Observation of pair tunneling and coherent destruction of tunneling in arrays of optical waveguides. Phys. Rev. A 94, 053853 (2016).

40. Tai, M. E. et al. Microscopy of the interacting Harper–Hofstadter model in the two-body limit. Nature 546, 519–523 (2017).

41. Olekhno, N. A. et al. Topological edge states of interacting photon pairs emulated in a topolectrical circuit. Nat Commun 11, 1436 (2020).

42. Léonard, J. et al. Realization of a fractional quantum Hall state with ultracold atoms. Nature 619, 495–499 (2023).

43. Azcona, P. M. & Downing, C. A. Doublons, topology and interactions in a one-dimensional lattice. Sci Rep 11, 12540 (2021).

44. Paul, B. & Mishra, T. Realizing nontrivial doublon formation using a quantum computer. Phys. Rev. B 110, L020302 (2024).

45. Su, W. P., Schrieffer, J. R. & Heeger, A. J. Solitons in Polyacetylene. Phys. Rev. Lett. 42, 1698–1701 (1979).

46. Malkova, N., Hromada, I., Wang, X., Bryant, G. & Chen, Z. Observation of optical Shockley-like surface states in photonic superlattices. Opt. Lett., OL 34, 1633–1635 (2009).

47. Rudner, M. S., Lindner, N. H., Berg, E. & Levin, M. Anomalous Edge States and the Bulk-Edge Correspondence for Periodically Driven Two-Dimensional Systems. Phys. Rev. X 3, 031005 (2013).

48. Maczewsky, L. J. et al. Fermionic time-reversal symmetry in a photonic topological insulator. Nat. Mater. 19, 855–860 (2020).

49. Drüeke, H., Meschede, M. & Bauer, D. Steering edge currents through a Floquet topological insulator. Phys. Rev. Research 5, 023056 (2023).

50. Szameit, A. & Nolte, S. Discrete optics in femtosecond-laser-written photonic structures. J. Phys. B: At. Mol. Opt. Phys. 43, 163001 (2010).